\documentclass[prb,showpacs,twocolumn]{revtex4}
\usepackage{graphicx,psfrag,amssymb,amsmath}
\begin{document}
\title{Unconventional density wave in CeCoIn$_5$?}
\author{Bal\'azs D\'ora}
\affiliation{The Abdus Salam ICTP, Strada Costiera 11, I-34014, Trieste, Italy}
\author{Kazumi Maki}
\affiliation{Department of Physics and Astronomy, University of Southern California, Los Angeles CA 90089-0484, USA}
\author{Attila Virosztek}
\affiliation{Department of Physics, Budapest University of Technology and Economics, H-1521 Budapest, Hungary}
\affiliation{Research Institute for Solid State Physics and Optics, P.O.Box
49, H-1525 Budapest, Hungary}
\author{Andr\'as V\'anyolos}
\affiliation{Department of Physics, Budapest University of Technology and Economics, H-1521 Budapest, Hungary}

\date{\today}

\begin{abstract}
Very recently large Nernst effect and Seebeck effect were observed above
the superconducting transition temperature $2.3$~K in a heavy fermion
superconductor CeCoIn$_5$. We shall interpret this large Nernst effect in
terms of unconventional density wave (UDW), which appears around
$T=18$~K. Also the temperature dependence of the Seebeck coefficient below
$T=18$~K is described in terms of UDW. Another hallmark for UDW is the
angular dependent magnetoresistance, which should be readily accessible
experimentally.
\end{abstract}


\maketitle
\section{Introduction}
The new heavy fermion superconductor CeCoIn$_5$ discovered recently has
attracted considerable attention\cite{petrovic}.
For example there are many parallels between CeCoIn$_5$ and high $T_c$
cuprate superconductors: the quasi-two dimensionality, d-wave
superconductivity and the appearance of superconductivity in the vicinity
of antiferromagnetic state\cite{moshovich,izawa,sidorov,paglione}.

\begin{figure}[h!]
\psfrag{x}[t][b][1][0]{$B$}
\psfrag{nfl}[][][1][0]{UDW}
\psfrag{fl}[t][b][1][0]{Fermi liquid}
\psfrag{sc}[][][1][0]{d-wave SC}
\psfrag{y}[b][t][1][90]{$T$}
\includegraphics[width=7cm,height=4cm]{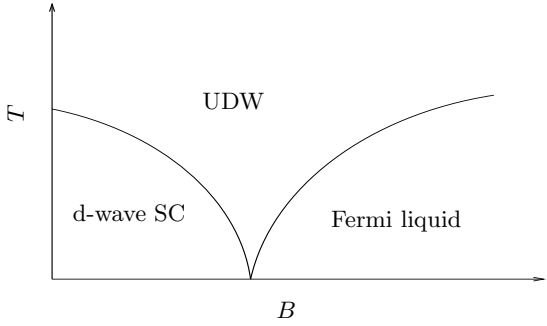}
\caption{The schematic magnetic field-temperature phase diagram of CeCoIn$_5$ after
Ref. \onlinecite{paglione}.}
\label{phase}
\end{figure}

Very recently Bel et al. in Ref. \onlinecite{bel} reported large thermoelectric power 
and Nernst
effect in CeCoIn$_5$ above the superconducting transition temperature. The
large thermoelectric effect is naturally understood in terms of the Kondo
lattice\cite{bel}. The schematic phase diagram of CeCoIn$_5$ is shown in Fig. 
\ref{phase}. From the Fermi liquid to UDW region, the resistivity 
changes\cite{bianchi} from $T^2$ 
to $T$.
In this Letter we want to show that both the Seebeck coefficient and the
Nernst effect below $T<18$~K are described in terms of unconventional
density wave (UDW).

UDW can be unconventional charge density wave or unconventional spin
density wave, though further experiments are needed to select one of
them. UDW is a kind of density wave with the quasiparticle energy gap 
$\Delta({\bf k})$, which has usually nodes on the Fermi surface. Here $\bf
k$ is the quasiparticle wave vector\cite{nagycikk}. Many people think now that the
pseudogap phase in high $T_c$ superconductors is UDW\cite{benfatto,nayak,krakko}. 
Also we have shown recently that the low temperature phase of
$\alpha$-(BEDT-TTF)$_2$KHg(SCN)$_4$ is UCDW\cite{admrprl,fujita,nernst}. In fact high 
$T_c$ cuprates in
the pseudogap regime\cite{wang1,capan} and the LTP in $\alpha$-(BEDT-TTF)$_2$KHg(SCN)$_4$
exhibit large negative Nernst effect\cite{choi}.

The quasiparticle energy in UDW in a magnetic field is quantized\cite{Ners1}. Then in
an electric field $\bf E$ within the conducting plane, the quasiparticle
orbits drift with the velocity ${\bf v}_D={\bf E\times B}/B^2$. This gives
rise to the transverse heat current ${\bf J}_{heat}=TS {\bf v}_D$, where
$T$ and $S$ are the temperature and the entropy associated with the
quasiparticles, respectively\cite{nernst}.

As to the Seebeck effect below $T<18$~K, we have to assume that the large
thermoelectric power around $T\sim 20$~K is due to the Kondo effect. We
assume also that the Kondo lattice instability is disrupted by the
appearance of UDW around $T=18$~K. The present model describes the $T$
linear electric resistance\cite{paglione,bianchi} observed for 
$T<10$~K$\sim T_c/2$.

\section{Quasiparticle spectrum}

The quasiparticle spectrum in UDW in CeCoIn$_5$ is given by 
\begin{equation}
E^\pm ({\bf k})=\pm\sqrt{v^2(k-k_F)^2+\Delta^2\cos^2(2\phi)}-\mu,
\end{equation}
where $v$, $\Delta$ and $\mu$ are the Fermi velocity, the maximum of the
energy gap and the chemical potential, respectively. Here we have assumed
d-wave DW as in high $T_c$ cuprates\cite{benfatto,nayak,krakko}. In the vicinity of the nodal points it
is convenient to replace $\Delta^2\cos^2(2\phi)$ by $v_\perp^2 k_\perp ^2$,
where $v_\perp/v=\Delta/E_F$\cite{chiao}. Now in a magnetic field tilted by an angle
$\theta$ from the c-axis, the energy spectrum becomes\cite{Ners1,nernst}
\begin{equation}
E_n^\pm=\pm\sqrt{\frac{2en}{m^*}|B\cos(\theta)|\Delta(T)}-\mu,
\end{equation}
where $m^*=\hbar k_F/v$ and $n=0$, $1$, $2$\dots.
The energy spectrum is very similar to the one of the Dirac particle in a
magnetic field\cite{weisskopf}. From the above quasiparticle spectrum, the electric
conductivity and Seebeck coefficient are obtained as
\begin{gather}
\sigma=\sigma_0+\sum_{n>0,\pm}\sigma_n\frac{1}{\exp(x_n^\pm)+1},\\
S_{xx}=\frac{\pi^2}{3e}T\frac{\partial}{\partial\mu}\ln(\sigma(\mu))
=\nonumber\\
\frac{\pi^2}{3e}T\sigma^{-1}\left\{\sum_{n>0,\pm}\frac{\partial\sigma_n}
{\partial\mu}\left(exp(x_n^\pm)+1\right)^{-1} \right.\nonumber \\
\left.\mp\frac{1}{T}\sum_{n>0,\pm}\sigma_n\exp(x_n^\pm)\left(\exp(x_n^\pm)+1\right)^{-2}\right\},
\label{seebeck}
\end{gather}
where $x_n^\pm=(\sqrt{2en\Delta(T)|B\cos(\theta)|/m^*}\pm\mu)/T$. Here we took the 
standard expression of the thermoeletric power\cite{abrikosov}. A more systematic 
analysis of the Kondo effect plus Landau quantization will be dealt with in a future 
paper.
Also we have assumed
$\partial\sigma_n/\partial\mu\sim\sigma_n/T_K$ with the Kondo temperature
$T_K\sim 20$~K, and this is expected to be the dominant contribution to the Seebeck 
coefficient.
The Nernst effect is given on the other hand by\cite{nernst} 
\begin{equation}
\alpha_{xy}=-\frac{S}{B\sigma},
\label{nernst}
\end{equation}
where the entropy $S$ reads as
\begin{gather}
S=\frac{g(0)e|B\cos(\theta)|}{m^*}\left[\ln(2)+\right.\nonumber\\
\left.\sum_{n>0,\pm}\left\{2\ln\left(2\cosh\left(\frac{x_n^\pm}
{2}\right)\right)-x_n^\pm\tanh\left(\frac{x_n^\pm}{2}\right)\right\}\right],
\end{gather}
where $g(0)$ is the density of states at the Fermi energy in the normal state.

\section{Comparison with experiments}

\begin{figure}[h!]
\psfrag{x}[t][b][1][0]{$B$(T)}
\psfrag{y}[b][t][1][0]{$S_{xx}$($\mu$V/K)}
\includegraphics[width=7cm,height=7cm]{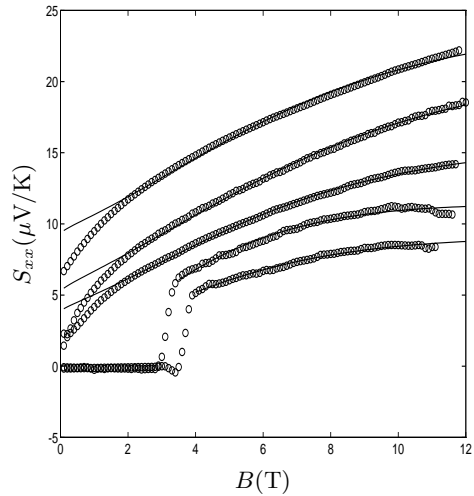}
\caption{The magnetic field dependence of the Seebeck coefficient is shown for
$T=1.3$~K, $1.65$~K, $2.5$~K, $3.5$~k and $4.8$~K from bottom to top. 
The circles denote the experimental data, the solid line is our fit.}
\label{seeb1}
\end{figure}

\begin{figure}[h!]
\psfrag{x}[t][b][1][0]{$B$(T)}
\psfrag{y}[b][t][1][0]{$S_{xx}$($\mu$V/K)}
\includegraphics[width=7cm,height=7cm]{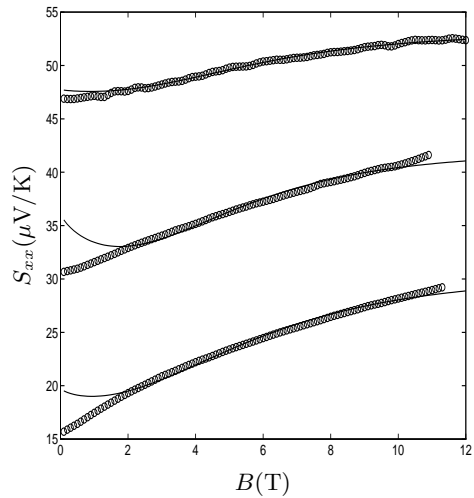}
\caption{The magnetic field dependence of the Seebeck coefficient is shown for
$T=7.3$~K, $10.5$~K and $15$~K from bottom to top.
The circles denote the experimental data, the solid line is our fit.}
\label{seeb2}
\end{figure}

\begin{figure}[h!]
\psfrag{x}[t][b][1][0]{$B$(T)}
\psfrag{y}[b][t][1][0]{$\alpha_{xy}$($\mu$V/K)}
\includegraphics[width=7cm,height=7cm]{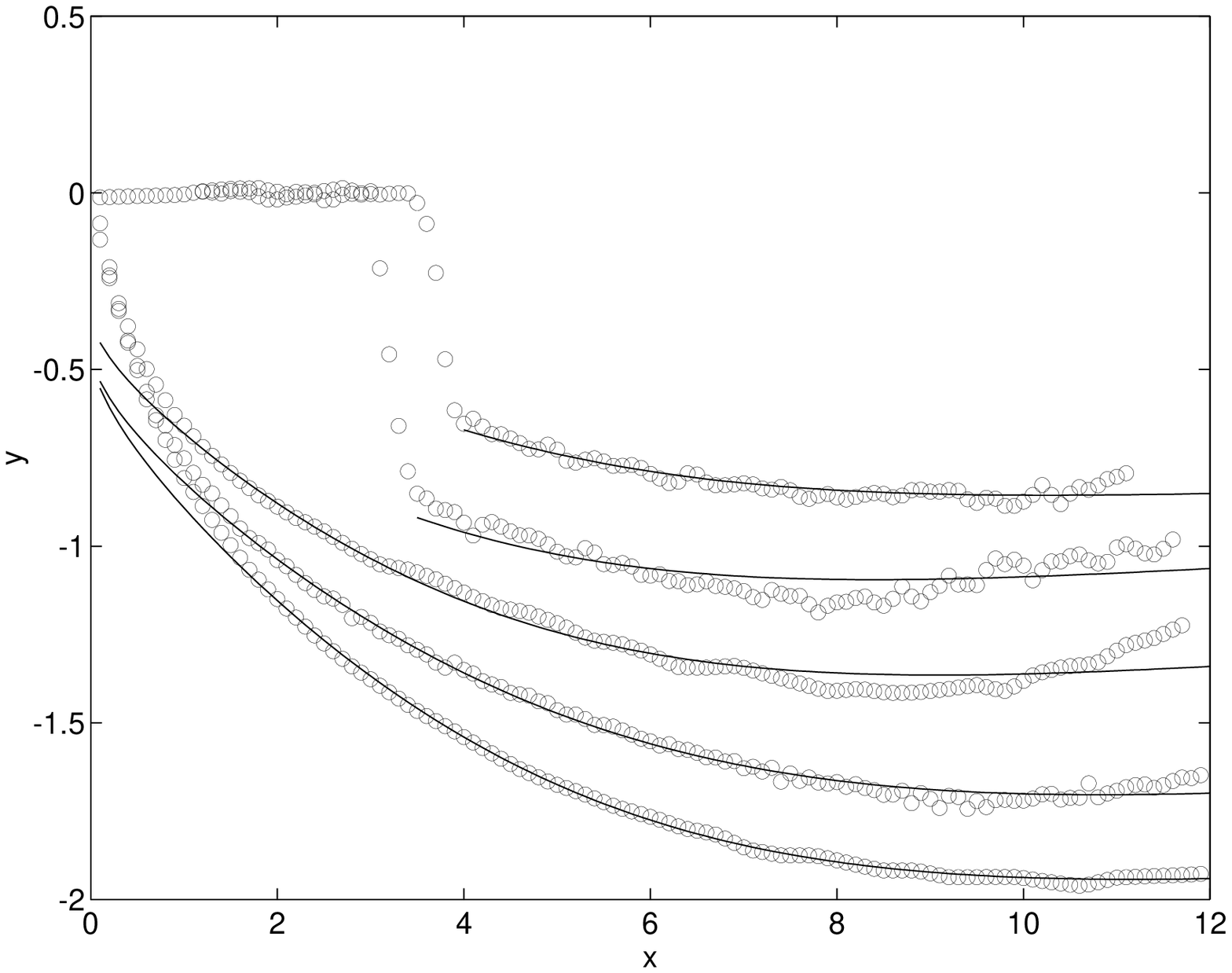}
\caption{The magnetic field dependence of the Nernst coefficient is plotted for
$T=1.3$~K, $1.65$~K, $2.5$~K, $3.5$~k and $4.8$~K from top to bottom. 
The circles denote the experimental data, the solid line is our fit.}
\label{ner1}
\end{figure}

\begin{figure}[h!]
\psfrag{x}[t][b][1][0]{$B$(T)}
\psfrag{y}[b][t][1][0]{$\alpha_{xy}$($\mu$V/K)}
\includegraphics[width=7cm,height=7cm]{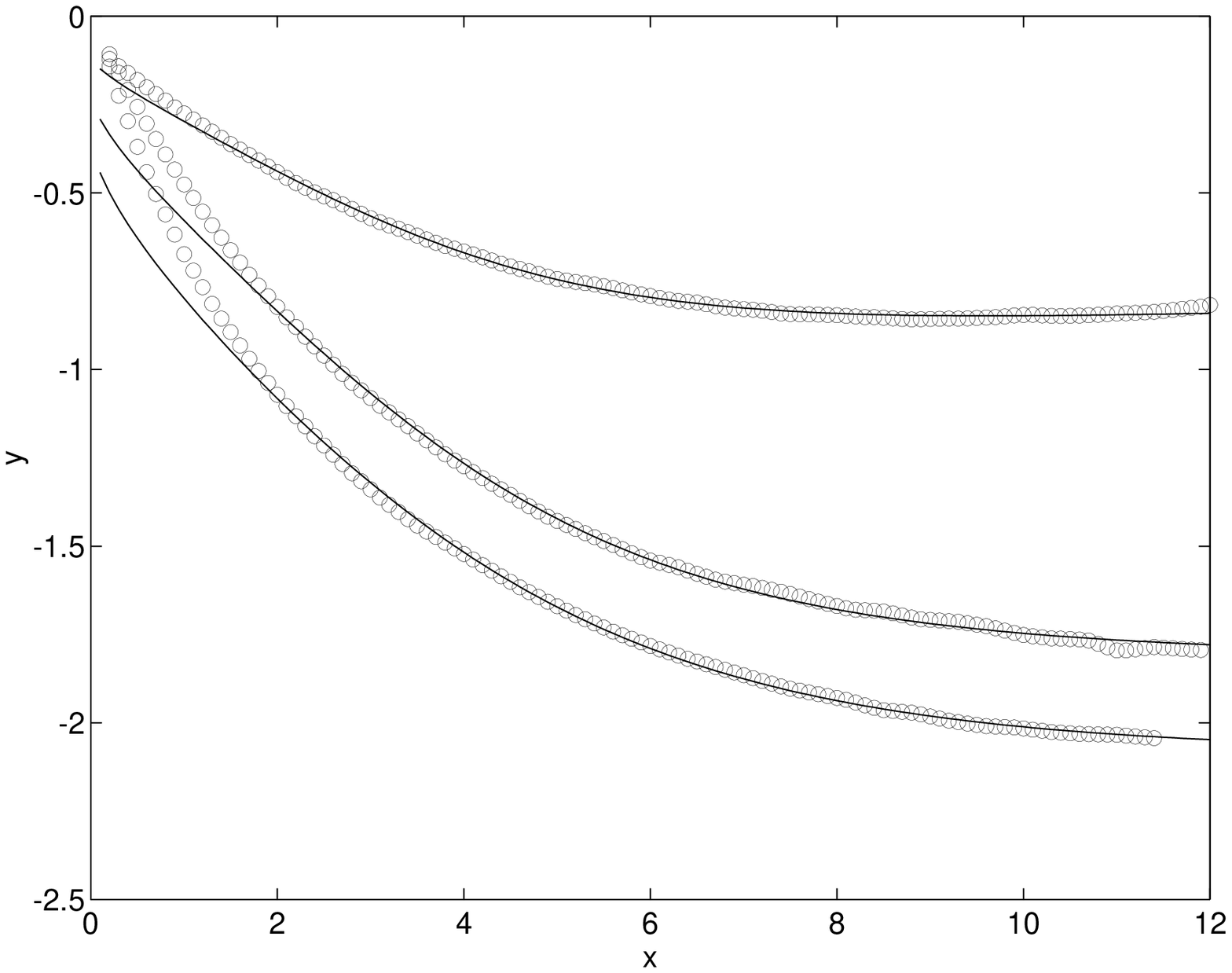}
\caption{The magnetic field dependence of the Nernst coefficient is shown for
$T=7.3$~K, $10.5$~K and $15$~K from bottom to top.
The circles denote the experimental data, the solid line is our fit.}
\label{ner2}
\end{figure}

\begin{figure}
\psfrag{x}[t][b][1][0]{$T$(K)}
\psfrag{y}[b][t][1][0]{$\sigma_n(T)/\sigma_n(T=10.5$~K)}
\includegraphics[width=7cm,height=7cm]{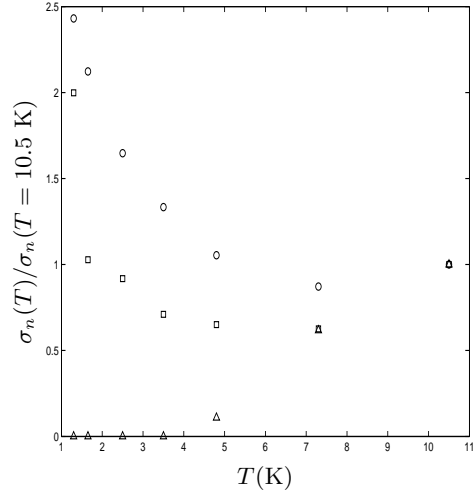}
\caption{The temperature dependence of $\sigma_n$, $n=0$ (circles), $1$ (triangles)
and $2$ (squares) is shown as extracted from previous fittings. Below $T=5$~K, 
$\sigma_0$ and $\sigma_2$ are proportional to $1/T$, $\sigma_1$ is finite, $T$ independent.}
\label{sigmas}
\end{figure}

\begin{figure}
\psfrag{x}[t][b][1][0]{$T$(K)}
\psfrag{y}[b][t][1][0]{$(\Delta(T)/m^*)^{1/4}$ (arbitrary units)}
\includegraphics[width=7cm,height=7cm]{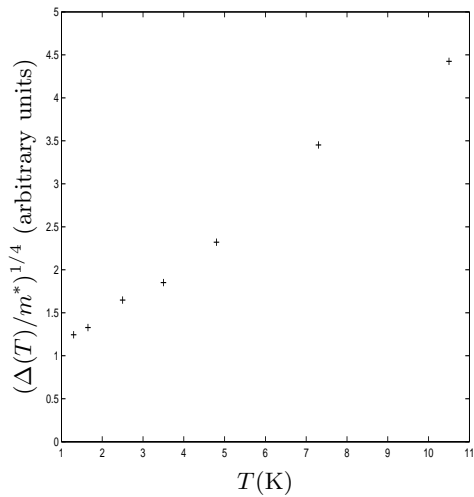}
\caption{The temperature dependence of $(\Delta(T)/m^*)^{1/4}$ is plotted. Assuming a 
constant $\Delta\sim 40$~K, $m^*$ varies as $1/(c+T)^4$, $c>0$ constant.}
\label{delta}
\end{figure}

We first show $S_{xx}$ and $\alpha_{xy}$ versus $B$  
in
Figs. \ref{seeb1}, \ref{seeb2}, \ref{ner1} and \ref{ner2}  for several
temperatures. The fittings to the data from Ref. \onlinecite{bel} appear to be 
excellent except for very small   
fields. But these deviations are expected, since we used only 3 Landau
levels ($n=0$, $1$ and $2$) in these fittings. Also the fittings are expected to break
down when the system becomes superconducting ($B<H_{c2}$ in Figs. \ref{seeb1} and \ref{ner1}).
From the fittings, the temperature dependence of the conductivities
($\sigma_n$) can be extracted. Their behaviour look somewhat strange, 
as seen in Fig. \ref{sigmas}.
First of all there is a clear break at $T\sim 5$~K, which may indicate some unknown
transition. Below this temperature, $\sigma_0$ and $\sigma_2$ varies as $1/T$, while
$\sigma_1$ turns out to be constant. The former is consistent with the 
observation\cite{bianchi}
$\rho=1/\sigma\sim T$.
Their ratio at $10.5$~K was found to be $\sigma_0/\sigma_2=0.02$ and 
$\sigma_1/\sigma_2=0.5$, which suggests that the resistivity is dominated by the $n=2$ 
Landau level, but $\sigma_1$ becomes also important with increasing temperature. Also  
the temperature dependence of $\Delta(T)/m^*$ is unusual. By 
assuming 
$\Delta(T)\sim 40$~K independent of temperature in this $T$ region, $v\sim 1/m^*$
increases with
increasing temperature as $(c+T)^4$, $c>0$ constant. Perhaps this 
can be the 
manifestation of quantum critical
point in CeCoIn$_5$. Therefore, in spite of somewhat unusual temperature dependence of 
physical quantities like $\sigma_n(T)$ and $v(T)$, the simple theoretical expressions 
for the Seebeck and Nernst coefficients (Eqs. \ref{seebeck} and \ref{nernst}) work 
very well to describe the magnetic field dependence of $S_{xx}$ and $\alpha_{xy}$. 

\section{Conclusion}

In summary we have analysed recent magnetothermopower data from CeCoIn$_5$
for $T<20$~K in terms of UDW, which appears around $18$~K. As stressed
elsewhere, the large negative Nernst effect is the hallmark of UDW\cite{mplb}. Indeed,
UDW provides us with excellent description of both Seebeck coefficient and
Nernst coefficient observed in CeCoIn$_5$.
This situation is very similar to what we encounter in 
$\alpha$-(BEDT-TTF)$_2$KHg(SCN)$_4$\cite{mplb} salt, (TMTSF)$_2$PF$_6$\cite{tmtsf} and in  high 
$T_c$ 
cuprates YBCO, LSCO and Bi2212\cite{capnernst}.


Also we recall, that the large Nernst effect observed in NbSe$_2$\cite{bel2}
indicates, that CDW in this material should be UCDW\cite{castroneto}. We expect also that the
giant Nernst effect will provide definitive signature of UDW in candidate
systems like the antiferromagnetic phase in URu$_2$Si$_2$\cite{bel3}, CeRhIn$_5$,
CeCu$_2$Si$_2$, UBe$_{13}$ and the glassy phase in $\kappa$-(ET)$_2$ salts.

\begin{acknowledgments}

We thank Kamran Behnia and Yuji Matsuda for providing us with their
experimental data prior to publication. We are grateful to R. Bel, Y. Matsuda,
S. Haas and H. Won for useful discussion on CeCoIn$_5$.
K. M. and B. D. acknowledge gratefully the hospitality of the Max Planck Institute for the
Physics of Complex Systems and Max Planck Institute for Chemical Physics of
Solids at Dresden, where some part of this work was done.
This work was supported by the Hungarian
Scientific Research Fund under grant numbers OTKA T046269, NDF45172 and
TS040878.
\end{acknowledgments}
\bibliographystyle{apsrev}
\bibliography{ceco}

\end{document}